\documentstyle[twocolumn,prc,aps,epsfig]{revtex}
\newcommand{\be}{\begin{eqnarray}}
\newcommand{\ee}{\end{eqnarray}}

 \newcommand{\gsim}{\mathrel{\hbox{\rlap{\lower.55ex \hbox {$\sim$}}
                   \kern-.3em \raise.4ex \hbox{$>$}}}}
\newcommand{\lsim}{\mathrel{\hbox{\rlap{\lower.55ex \hbox {$\sim$}}
                   \kern-.3em \raise.4ex \hbox{$<$}}}}

\def\roughly#1{\mathrel{\raise.3ex\hbox{$#1$\kern-.75em%
\lower1ex\hbox{$\sim$}}}}
\def\lsim{\roughly<}
\def\gsim{\roughly>}

\newcommand{\sigpompom}{\mbox{$\sigma_{{\cal P}{\cal P}}^{tot}$}}

\newcommand{\eq}{\begin{equation}}
\newcommand{\eqx}{\end{equation}}
\newcommand{\eqn}{\begin{eqnarray}}
\newcommand{\eqnx}{\end{eqnarray}}

\begin{document}

\twocolumn[\hsize\textwidth\columnwidth\hsize\csname @twocolumnfalse\endcsname

\title {Semiclassical Double-Pomeron Production of Glueballs and
 $\eta'$ }
\author {Edward Shuryak and Ismail Zahed}
\address{Department of Physics and Astronomy, State University
of New York, Stony Brook, NY 11794}

\date{\today}
\maketitle
\begin{abstract}
A semiclassical theory of high energy scattering 
based on interrupted tunneling (instantons) or
QCD sphaleron production has been recently developed to describe
the growing hadronic cross section and properties of the soft Pomeron.
In this work we address double-pomeron processes in this
framework for the first time. We specifically
derive the cross section for central production of parity even and odd
clusters, scalar and pseudoscalar glueballs, and 
$\eta'$ in  parton-parton scattering at high energy. We show that
the specific dependence of the production cross section
on all its kinematical variables compares favorably with the 
UA8 data on inclusive cluster production, as well as the WA102 data
on exclusive central production of scalar
glueball and $\eta'$, in double-pomeron exchange $pp$ scattering.
The magnitude of the cross section and its dependece on kinematic variables
is  correct, explaining in
particular a large deviation from the Pomeron factorization at 
cluster masses in the range $M_X<8$ GeV reported by UA8.
\end{abstract}
\vspace{0.1in}
]
\begin{narrowtext}
\newpage

\section{Introduction}\label{intro}

Semi-classical tunneling in the QCD vacuum, described by
instantons, is traditionally studied in relation with the QCD
vacuum properties such as chiral symmetry breaking and hadronic
spectroscopy, see review \cite{SS_98}.

More recently a number of authors \cite{sz01,KKL} have suggested that
the semiclassical physics based on instantons and QCD
sphalerons significantly contributes  to semi-hard scattering in
QCD, in particular to the parameters of the so called ``soft Pomeron''.
The specific behavior of hadronic cross sections
 at high energy, i.e. their growth with energy
$\sigma \sim s^{0.08}$ is related to the Pomeron trajectory intercept
at $t=0$. The semiclassical theory relates the small power of $0.08$ 
to the barrier-suppressed probability of tunneling in the QCD vacuum. 
Also, the small Pomeron size $\alpha'=1/(2 \, {\rm GeV})^2=(0.1 \,
{\rm fm})^2$ was found to be related in~\cite{sz01,KKL} to the small 
instanton size $\rho=1/3$ fm. As unexpected bonus, it was found that
the semiclassical scattering cannot produce an {\em odderon},
essentially due to the inherent $SU(2)$ color nature of the
semiclassical  fields. Recently, the same semiclassical reasoning was 
applied to the saturation problem at HERA~\cite{schrempp}.

The semiclassical approach to semi-hard processes is distinct
from many QCD models in a number of ways. It describes field
excitations from the under-the-barrier part of vacuum wave 
function, becoming on-the-barrier states referred to as QCD sphalerons.
They are specific topological clusters made of purely magnetic glue~\cite{OCS}.
As quantum-mechanical and semi-classical
arguments show,  it is the most natural excitation of the glue
from {under the barrier}.  When produced they explode~\cite{OCS,jsz02},
creating on the way many light quark pairs~\cite{SZ}.

The corresponding contributions to the
soft Pomeron can be viewed as a ladder-type
diagrams, similar to the perturbative BFKL ones but with
different rungs~\cite{KKL,sz01}. The Lipatov vertex -- 
2 virtual gluons fusing into  one physical gluon --
is substituted by a new vertex with a tunneling path ending
at the unitarity cut at the ``turning state'' -the  topological clusters.
In this work we  focus on only one cluster production as illustrated
in Fig. 1. 

In contrast to deep-inelastic scattering,
high energy  hadronic collisions in the semi-hard regime have
no large scale $Q^2$,
and so the produced clusters have masses and sizes that are determined
by the typical  size of the instantons in the QCD vacuum. This leads to
a mass of order 3 GeV for a size of order $1/3$ fm, as mentioned above.

A significant amount of clustering
in $pp$ collisions has been known for a very long time~\cite{clans}, where it was also
pointed out that those clusters have on average larger mass and
multiplicity  in comparison to the clusters produced in
$e^+e^-$ annihilation. Unfortunately, a study of
these clusters and their identification is still not done. In general,
from the analysis of  secondaries in $pp$ collisions it is hard to tell
which clusters are sphaleron-related and produced promptly, and which
are simply products of the string fragmentation, a final state interaction
unrelated to the underlying dynamics responsible for the
cross section. In ordinary inclusive $pp$ collisions only a bulk
statistical analysis can be performed.

\begin{figure}[ht]
\begin{center}
\mbox{\epsfig{file=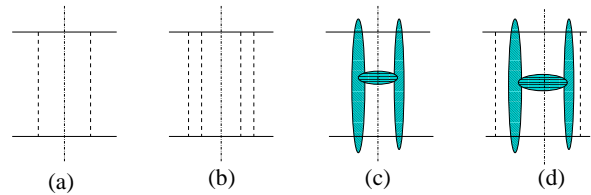,width=8cm}}
\end{center}
\caption[]{
Schematic
diagrams for the cross sections of different processes associated with
 high energy collusion of two quarks, shown by horizontal solid
lines. The vertical dash-dotted lines are unitarity cuts, they
separate the amplitude from its complex conjugate. (a) Low-Nussinov or
single-gluon exchange, leading to inelastic collisions due to color
exchange;
(b) Low-Nussinov  cross section, with no color exchange;
(c) instanton-induced inelastic collision with color transfer and
prompt cluster production, (d) combined instanton-gluon process
leading to double-pomeron like  events with a cluster.
}
\label{fig:fig_tHooft_1storder}
\end{figure}

That is why in the present work we focus on double-pomeron  scattering,
or processes in which there are two large {\em rapidity gaps} separating
the colliding two protons with a single  cluster  produced at mid-rapidity.
In this case there is no place for color strings and their
fragmentation, as all object involved
 are colorless. So, if our assumption about dominance
of the topological clusters is correct, we should be able to describe
the double-pomeron data solely from the semiclassical theory. The
answer is yes, as we will detail below.

Important experimental findings on $inclusive$
double diffraction were recently
reported by the UA8 collaboration~\cite{UA8}, based on 
its 1989 data data sample 
at CERN $S\bar p pS$ collider. We note in this
work that the reported data display a wide maximum around cluster
masses of order few GeV, with a cross section that is an order of
magnitude larger than the one predicted by Pomeron factorization.
Interestingly enough, the clusters with mass less than 5 GeV
decay isotropically in their rest frame. Unfortunately, the
UA2 detector used in this work was a simple calorimeter with poor mass
resolution with sigma about 2 GeV, which prohibited  from seeing mass
structure in our pomeron-pomeron cross section.

The WA102 collaboration at CERN carried a fixed target $pp$
experiment at  $\sqrt{s}= 28$ GeV, focusing on the double-pomeron
exclusive production into few  hadron states. This experiment was the 
first to discover a strong dependence of the cross section on the
azimuthal angle between the momenta transfered to
two protons, a feature that was not expected from standard Pomeron
phenomenology. This result inspired some phenomenological works
\cite{close,EK,koch} pointing a possible  analogy between the Pomeron and
vector particles. Close and his collaborators have even suggested
to use this azimuthal distribution as a glueball  filter, selecting
the hadronic  states  which peak at small
difference in transverse momentum $dP_T$ of the  protons. In
particular, the production of scalars and tensors
such as $f_0(980)$, $f_0(1500)$, $f_J(1710)$,
and $f_2(1900)$ was found to be considerably enhanced at small $dP_T$,
while the production of pseudoscalars such as $\eta$, $\eta'$ was
found to be peaked at mutually orthogonal momentum transfers of the
protons. In our approach the produced QCD  sphalerons can be
regarded as precursors of glueballs or pseudoscalar $\eta'$ strongly
coupled to glue. In the double-pomeron process the sphaleron
production dwarfs the instanton-antiinstanton process~\cite{koch}
by 2 orders of magnitude, and contrary to the latter it triggers
a rise in the cross section.

The paper is organized as follows: In section 2 we recall the general
expression for  the total QCD cross section in the eikonal approximation, and check
its perturbative (Low-Nussinov) limit. In section 3 we analyze the generic
form of the double-pomeron cross section and show its
direct relationship with the inelastic cross section through
the Pomeron.  In section 4, we discuss the double-pomeron
inclusive UA8 results in light of our results. In section 5, we derive
explicit results for the even/odd double-pomeron gluon production.
In section 6 we use the scale and U(1) anomaly to derive
the double-pomeron cross sections for isosinglet production.
In section 7, we compare our results to the
CERN WA102 results for the reported glueball and $\eta'$ states. Our conclusions
are in section 8.

\section{Total Cross Section}

In this section and to help streamline the definitions,
we quote the general result for the instanton-induced contribution
to the total cross section, and check its perturbative limit.

\subsection{General Result}

The generic form of the total cross section for parton-parton scattering
through a generic gauge configuration in QCD reads at large $\sqrt{s}$
\cite{sz01}

\be
&&\sigma \approx \frac{\rm Im}{VT}\,\sum_{CD}\,\frac{1}{(2\pi)^6}
\,\int dq_{1+}\,dq_{1\perp}\,dq_{2-}\,dq_{2\perp}\nonumber\\
&&\times \int [dA][dA']\,e^{iS(A)-iS(A')+
iS(A,A')}\,\nonumber\\
&&\times\int dx_-dx_\perp dy_+dy_\perp \,
e^{\frac i2 q_{1+}x_--iq_{1\perp}
x_\perp +\frac i2 q_{2-}y_+ -iq_{2\perp}y_\perp}\nonumber\\
&&\times\left( {\bf W}_- (\infty, x_-, x_\perp) -{\bf 1}\right)_{AC}
\left( {\bf W}_+ (y_+, \infty, y_\perp) -{\bf 1}\right)_{BD}\nonumber\\
&&\times\int dx'_-dx'_\perp dy'_+dy'_\perp \,e^{\frac i2 q_{1+}x'_--iq_{1\perp}
x'_\perp +\frac i2 q_{2-}y'_+ -iq_{2\perp}y'_\perp}\nonumber\\
&&\times\left( {\bf W}_- (\infty, x'_-, x'_\perp) -{\bf 1}\right)^*_{AC}
\left( {\bf W}_+ (y'_+, \infty, y'_\perp) -{\bf 1}\right)^*_{BD}\,\,.
\nonumber\\
\label{A1}
\ee
which is the imaginary part of a retarded 4-point correlation
function in Minkowski space. The line integrals are along the
light cone with

\be
&&{\bf W}_{-} (\infty, x_{-}, x_\perp ) = \nonumber\\
&&{\bf P}_c \,{\rm exp} \left(-\frac {ig}2\int_{-\infty}^{+\infty}
dx_+' \,A_- (x_+', x_{-}, x_\perp )\right)\,\,.
\label{A2}
\ee
and

\be
&&{\bf W}_{+} (x_{1+}, \infty, x_\perp ) = \nonumber\\
&&{\bf P}_c \,{\rm exp} \left(-\frac {ig}2\int_{-\infty}^{+\infty}
dx_-' \,A_+ (x_{+}, x_{-}', x_\perp )\right)\,\,.
\label{A3}
\ee

\subsection{Low-Nussinov Limit}

In perturbation theory, the line integrals can be expanded to give
in the lowest perturbative order
\be
{\bf W}_+ -{\bf 1}\approx
-\frac {ig}2\int_{-\infty}^{+\infty} dx_-' \,A_+ (x_{+}, x_{-}', x_\perp)
\label{A4}
\ee
and similarly for ${\bf W}_-$. The correlator of two vector potentials
is the gluon propagator, so the expression corresponds to a diagram
shown in Fig.1(a). Inserting (\ref{A4}) into
(\ref{A1}) we obtain to leading order in perturbation theory

\be
\sigma \approx &&\left(\frac{\alpha_s}{\pi}\right)^2\,
\left(\frac{T^e}2\frac{T^f}2\right)_{AA}
\left(\frac{T^e}2\frac{T^f}2\right)_{BB}\,
\int\,\frac{dq_\perp}{\pi^2}\, \nonumber\\&&\times
\left|
\int db_\perp\,e^{-iq_\perp b_\perp}\, \int d\alpha\,d\beta\,
\frac{v_+\cdot v_-}{(v_+\alpha-v_-\beta)^2-b_\perp^2+i0}
\right|^2\nonumber\\
\label{A5}
\ee
where $v_\pm$ are the 4-velocities on the light-cone with
proper-time extent $T$. The result (\ref{A5}) is

\be
\sigma \approx &&\left(\frac{4\alpha_s}{\pi}\right)^2\,
\left(1-\frac1{N_c^2}\right)\,\nonumber\\&&
\times \int {dq_\perp}\,
\left|
\int db_\perp\,e^{-iq_\perp b_\perp}\, {\rm ln}\,\left(\frac T{b_\perp}\right)
\right|^2\,\,,\nonumber\\
\label{A6}
\ee
after proper color tracing, in agreement with the perturbative
result of Low-Nussinov~\cite{LOW}.

\section{Double-pomeron Cross Section}

The exclusive cross section for the double-pomeron parton-parton
scattering $qq\rightarrow qqX$ where $X$ stands
for a centrally produced color singlet cluster as
diagrammatically depicted in Fig. 1 (d). The
vertical ellipses indicate instantons and the dashed line is an additional
perturbative gluon that is needed to enforce overall color neutrality for
the state $X$ emitted in the central region. The thin vertical lines
in all diagrams denotes the unitarity cut. The corresponding exclusive
cross section reads

\be
&&\sigma \approx \frac {\rm Im}{VT}\,\frac{1}{(2\pi)^6}
\,\int dq_{1+}\,dq_{1\perp}\,dq_{2-}\,dq_{2\perp}\nonumber\\
&&\times \int [dA][dA']\,e^{iS(A)-iS(A')+
iS(A,A')}\,\nonumber\\
&&\times\int dx_-dx_\perp dy_+dy_\perp \,
e^{\frac i2 q_{1+}x_--iq_{1\perp}
x_\perp +\frac i2 q_{2-}y_+ -iq_{2\perp}y_\perp}\nonumber\\
&&\times\left( \frac{-ig}2\int_{-\infty}^{+\infty}\,dz_+
{A}_-(z_+,x_-, x_\perp)\,
{\bf W}_- (\infty, x_-, x_\perp)\right)_{AA}\nonumber\\
&&\times\left( \frac{-ig}2\int_{-\infty}^{+\infty}\,dz_-
{A}_-(y_+,z_-, y_\perp)\,
{\bf W}_+ (y_+, \infty, y_\perp) \right)_{BB}\nonumber\\
&&\times\int dx'_-dx'_\perp dy'_+dy'_\perp \,e^{\frac i2 q_{1+}x'_--iq_{1\perp}
x'_\perp +\frac i2 q_{2-}y'_+ -iq_{2\perp}y'_\perp}\nonumber\\
&&\times\left( \frac{-ig}2\int_{-\infty}^{+\infty}\,dz'_+
{A'}_-(z'_+,x'_-, x'_\perp)\,
{\bf W}_- (\infty, x'_-, x'_\perp)\right)^*_{AA}\nonumber\\
&&\times\left(\frac{-ig}2\int_{-\infty}^{+\infty}\,dz'_-
{A'}_-(y'_+,z'_-, y'_\perp)\,
{\bf W}_+ (y'_+, \infty, y'_\perp)\right)^*_{BB}\,\,.
\nonumber\\
\label{A7}
\ee
The leading order contribution to (\ref{A7}) stems from the
perturbative part of the gluon exchanged in Fig. 1 around the
semiclassical background attached to the eikonal partons. The
former increases with the proper-length of the eikonalized
trajectories spanned by the incoming partons. The result is

\be
&&\sigma \approx \frac {\rm Im}{VT}\frac{1}{(2\pi)^6}
\,\int dq_{1+}\,dq_{1\perp}\,dq_{2-}\,dq_{2\perp}\nonumber\\
&&\times \int [dA][dA']\,e^{iS(A)-iS(A')+
iS(A,A')}\,\nonumber\\
&&\times\int dx_-dx_\perp dy_+dy_\perp \,
e^{\frac i2 q_{1+}x_--iq_{1\perp}
x_\perp +\frac i2 q_{2-}y_+ -iq_{2\perp}y_\perp}\nonumber\\
&&\times 2\alpha_s\,{\rm ln}\left(\frac{T}{|x_\perp-y_\perp|}\right)
\nonumber\\
&&\times\left( \frac {T^a}2 {\bf W}_- (\infty, x_-, x_\perp)\right)_{AA}
\left(\frac{T^a}2 {\bf W}_+ (y_+, \infty, y_\perp)\right)_{BB}\nonumber\\
&&\times\int dx'_-dx'_\perp dy'_+dy'_\perp \,e^{\frac i2 q_{1+}x'_--iq_{1\perp}
x'_\perp +\frac i2 q_{2-}y'_+ -iq_{2\perp}y'_\perp}\nonumber\\
&&\times 2\alpha_s\,{\rm
ln}\left(\frac{T}{|x'_\perp-y'_\perp|}\right)\nonumber\\
&&\times\left( \frac {T^b}2{\bf W}_- (\infty, x'_-, x'_\perp) \right)^*_{AA}
\left( \frac {T^b}2 {\bf W}_+ (y'_+, \infty, y'_\perp) \right)^*_{BB}\,\,.
\label{A8}
\ee
The non-perturbative parts of the gluon propagator
in Feynman background gauge have been dropped.
The perturbative part dominates through
its logarithmic growth at large proper time $T$.

\subsection{Details}

The result (\ref{A8}) is general and holds in Minkowski space.
Following our previous arguments we assess the imaginary part
by continuing to Euclidean space and saturating the double
functional integral with singular gauge configurations which
are sphaleron-like. The result can be considerably simplified if
we note that the 2-dimensional Coulombic law probes transverse
distances $|x_\perp-y_\perp|$ of the order of the sphaleron
size $\rho$, while the Euclidean time $T$ is of the order of
the inverse sphaleron mass $1/M_s$. As a result and modulo
color factors, the 2-dimensional Coulomb contribution brings
about an overall factor of

\be
4\alpha_s^2\,{\rm ln}^2(\rho\,M_s) = 4\alpha_s^2\,
{\rm ln}^2\left(\frac{4\alpha_s}{3\pi}\right)\,\,.
\label{A9}
\ee
The color factors can be unwound by using

\be
{\bf W}={\rm cos}\,\alpha-i{\bf R}^{ai}\tau^a\,{\bf n}^i\,{\rm sin}\,\alpha
\label{A10}
\ee
where the singular gauge configuration is given by $\alpha$
\footnote{Since the singular gauge configuration asymptotes
the instanton profile, it is sufficient to use the
instanton/antiinstanton profile in the form factors.}.
Specifically, the contribution

\be
&&-i{\bf R}^{ai}{\bf n}^i\left(\frac{T^e}2\tau^a\right)_{AA}{\rm sin}\,\alpha
\nonumber\\
\times
&&-i{\bf R}^{bj}\underline{{\bf n}}^j
\left(\frac{T^e}2\tau^b\right)_{BB}{\rm sin}\,\underline{\alpha}
\nonumber\\
\times
&&+i{\bf R'}^{a'i'}{\bf n}^{i'}\left(\frac{T^f}2\tau^{a'}\right)^*_{AA}{\rm sin}\,\alpha'
\nonumber\\
\times
&&+i{\bf R'}^{b'j'}\underline{{\bf n}}^{j'}
\left(\frac{T^f}2\tau^{b'}\right)^*_{BB}{\rm sin}\,\underline{\alpha'}
\label{A11}
\ee
where the $T^e$'s are SU(N$_c$)generators with $e=1,..,(N_c^2-1)$
and $\tau^a$'s are SU(2) generators with $a=1,2,3$, can be simplified
using the color averaging relation

\be
\left(\frac{T^e}2\tau^a\right)_{AA}\equiv \frac 1{N_c}\,{\rm Tr}
\left(\frac{T^e}2\tau^a\right) =\frac{{\bf 1}^{ea}}{N_c}\,\,.
\label{A12}
\ee
Thus, (\ref{A11}) becomes

\be
\frac 1{N_c^4}\,{\bf n}\cdot\underline{\bf n}\,
{\bf n'}\cdot\underline{\bf n'}\,
{\rm sin}\,\alpha\,{\rm sin}\,\underline{\alpha}\,
{\rm sin}\,\alpha'\,{\rm sin}\,\underline{\alpha'}
\label{A13}
\ee

\subsection{Inclusive Double-Pomeron Cross Section}

Inserting (\ref{A9}-\ref{A13}) into (\ref{A8}) and following the
steps given in \cite{sz01}, we obtain for the total singlet cross
section

\be
\sigma (s) \approx &&{\bf C}_S\,\pi\,\rho^2\,{\rm ln}\,s\,
\int\,dq_{1\perp}\,dq_{2\perp}\,
\,{\bf K} (q_{1\perp} , q_{2\perp} ) \nonumber\\
&&\times\,\int^\infty_{(q_{1\perp}+q_{2\perp})^2}\, dM^2\,\,\sigma_S (M)\,\,.
\label{A14}
\ee
The constant ${\bf C}_S$ is equal to

\be
{\bf C}_S= \frac 1{(2\pi)^8}\,
\frac{64}{5N_c^2}\,\alpha_s^2\,
{\rm ln}^2\left(\frac{4\alpha_s}{3\pi}\right)\,\,.
\label{A15}
\ee

The partial cross section $\sigma_S (Q)$ is the same
as the one encountered in the sphaleron-like production.
To exponential accuracy

\be
\sigma_S (Q) = {\rm Im} \int\, dT\, e^{QT-{\bf S}(T)}\approx \kappa\,
e^{\frac{4\pi}\alpha\,({\bf F}(Q)-{\bf F}(M_s))}\,\,,
\label{A16}
\ee
where the holy grail function ${\bf F}(Q)$ was evaluated in
\cite{DP94,jsz02} using singular gauge configurations. In the
approximation

\be
\sigma_S (Q) \approx \kappa\,\,\delta (Q^2-M_s^2)
\label{A17}
\ee
the singlet cross section simplifies to

\be
\sigma  (s)\approx &&{\bf C}_S\,\pi\rho^2\,\kappa\,{\rm ln \,s}\,\,
\int\,dq_{1\perp}\,dq_{2\perp}\,
\,{\bf K} (q_{1\perp} , q_{2\perp})\,\,,
\label{A18}
\ee
which is analogous to the inelastic (Pomeron induced) cross section
derived in~\cite{sz01} with the substitution ${\bf C}_S\rightarrow {\bf
C}$ where

\be
{\bf C} = \frac 1{(2\pi)^8}\,\frac{64}{15}
%\frac{\Delta (0)}{\kappa}\left(\int dq_\perp{\bf J}^2(q_\perp})\right)^{-2}
\label{A19}
\ee
%where $\Delta(0)$ is the `Pomeron' intercept in our approach.
Note that the ratio of the semiclassical double-pomeron cross
section to the semiclassical inelastic  cross section, as given by
diagrams  Fig. 1d and 1c respectively,
is independent of the detailed dynamics, and involves mostly
color factors resulting from the singlet projection through
the extra gluon exchanged

\be
\frac{\sigma_{DD}}{\sigma_{IN}} \approx \frac{{\bf C}_S}{\bf C} =
\frac 3{N_c^2}\, \alpha_s^2\,{\rm
ln}^2\left(\frac{4\alpha_s}{3\pi}\right)\approx 0.125\,\,.
\label{A19x}
\ee

\subsection{The instanton-induced form factor}

The form factor  is
\be
{\bf K} (q_{1\perp} , q_{2\perp}) =
|{\bf J} (q_{1\perp}) \cdot {\bf J} (q_{2\perp})
+ {\bf J} (q_{1\perp})\times {\bf J} (q_{2\perp}) |^2
\,\,,
\label{A20}
\ee
with

\be
{\bf J} (q_{\perp}) = \int dx_3 \,dx_\perp \, e^{-iq_\perp x} \,
\frac{x_\perp}{|x|}\,{\rm sin} \left( \frac {\pi\,
|x|}{\sqrt{x^2+\rho_0^2}}\right)\,\,.
\label{A21}
\ee
which is purely imaginary,
\be
{\bf J} (q_{\perp}) =&& -i \frac{\hat{q}_\perp}{\sqrt{q_\perp}}
\,\,\int_0^\infty\,dx\,J_{3/2} (q_\perp x)\,\nonumber\\
&&\times\,\left( (2\,\pi x)^{3/2}\,{\rm sin}\left(\frac {\pi\,
|x|}{\sqrt{x^2+\rho_0^2}}\right)\right)\,\,.
\label{A22}
\ee
Here $J_{3/2}$ is a half-integer Bessel function. In the weak-field
limit the instanton contributes a term ${}^3\sqrt{x}/x^2\approx 1/\sqrt{x}$
that causes the instanton-induced form factor to diverge. This
divergence is analogous to the one encountered in $QQ\rightarrow QQ$.
Apart from the unphysical (perturbative) singularity at small $q_\perp$,
the instanton-induced form factor can be parameterized by a simple
exponential
\be
\label{A23}
{\bf J} (q_\perp) \approx -i\,{\hat{q}}_\perp\,50\, e^{-1.3\, q_\perp\, \rho_0} \,\,,
\ee
The divergence at
small $q_\perp$ can be removed by subtracting the tail of the instanton
through the substitution

\be
\frac {\pi\,
|x|}{\sqrt{x^2+\rho_0^2}} \rightarrow \pi\left( \frac {
|x|}{\sqrt{x^2+\rho_0^2}}-1\right)\,\,e^{-{{\bf a}\,|x|/\rho_0}}\,\,,
\label{A88}
\ee
which amounts to a different renormalization of the charge.
This will be understood throughout. Note that the subtracted
form factor vanishes at small $q_\perp$.

\section{Inclusive Double-Pomeron: UA8}

The UA8 experiment at CERN studied the reaction
${\overline p} p \rightarrow {\overline p} X p$
where $X$ is a set of hadrons at mid-rapidity. There are two
sets of data: one in which both nucleons were detected (AND) and one
in which only one nucleon was detected (OR). Since the two triggers
are different, the two data sets were measured at different kinematics.
UA8 used the following model-dependent parametrization of their measured
differential cross section

\be
&&{{d^6\sigma_{DPE}}\over{d\xi_1 d\xi_2 dt_1 dt_2 d\phi_1 d\phi_2}} \, \, = \nonumber\\&&
F_{{\cal P}/p}(t_1, \xi_1) \cdot F_{{\cal P}/p}(t_2, \xi_2) \cdot
\sigpompom (M_X).
\label{eq:pompomdist}
\ee
where the variables ($\xi_i, t_i, \phi_i$) describe the
fraction of the longitudinal momentum, momentum transfer squared
and its azimuthal direction for each Pomeron.  All the parameters
are uniquely given by the measured parameters of the outgoing
$p, \bar p$. The Pomeron flux factor or structure function is defined as

\be
F_{{\cal P}/p}(t, \xi)= K \,|{\bf F}_1(t)|^2 \, e^{bt} \, \xi^{1-2\alpha (t)}
\ee
where $|{\bf F}_1(t)|^2$ is the so called Donnachie-Landshoff~\cite{dl} nucleon form factor
\be
{\bf F}_1(t)={{4 m_p^2 - 2.8\,t} \over {4 m_p^2 - t}}\,\,
{1 \over (1-t/0.71)^2}\,\,.
\ee
The parameters were defined from single-pomeron data with 
$b=1.08 \pm 0.2 \, {\rm GeV}^{-2}$, and
nonlinear Pomeron trajectory 
\be
\alpha(t) \, = &&\, 1 + \epsilon + \alpha' t + \alpha'' t^2
\,\nonumber\\
 = &&\, 1.035 + 0.165 t + 0.059 t^2\,\,.
\label{eq:traj}
\ee
The parameter $K=0.74/{\rm GeV}^2$ was not measured and was set from
the Donnachie-Landshoff fit. The specific parametrizations
(\ref{pompomdist}-{traj}) were used to set up the acceptances and 
so on. However, in the UA8 paper to be discussed below,
it was pointed out that the difference between the AND and OR data
sets may suggest that the above parametrization with factorizable
flux factors maybe oversimplified.

The uncertainties related to the empirical extrapolation from the
covered to the full kinematical range notwithstanding, the UA8 data
show a striking and an unexpected shape and magnitude
\footnote{The extracted cross section is based on the value of $K$
quoted above.} for the pomeron-pomeron
cross section $\sigpompom(M_X)$ shown in Fig.~\ref{fig:sigpp}.
Only at the central cluster mass $M_X>10$ GeV was the cross section
small $\approx 0.1$ mb, and  more or less in agreement
with standard Pomeron calculus (more specifically with the
Pomeron factorization appended by some Reggeon contributions
decreasing with $M_X$). At smaller masses $M_X<10$ GeV the
observed cross section is an order of magnitude larger than
what is expected from factorization. This 
was neither predicted prior to the experiment, nor explained
(to our knowledge) after the experiment. Another crucial finding is
that the low-mass clusters decay isotropically (in their rest frame).
\begin{figure}
\begin{center}
\mbox{\epsfig{file=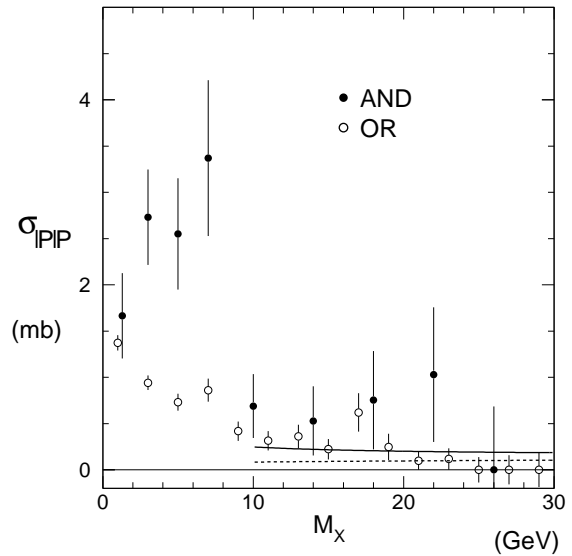,width=8cm}}
\end{center}
\caption[]{
Mass dependence of the Pomeron-Pomeron total cross section \sigpompom ,
derived from the AND  and OR triggered data, respectively.
The dashed curve is the factorization prediction (independent of $K$)
for the Pomeron-exchange component of \sigpompom .
The solid line is the fit to the OR points of a regge-exchange
term, $1/(M_X^2)^{0.32}$, added to this Pomeron-exchange term.
}
\label{fig:sigpp}
\end{figure}

Now, let us see to what extent our semiclassical description
is able to describe these observations. We start with a qualitative argument
of why the factorization does not work in our approach. This happens
because instead of a single universal Pomeron pole we view
high energy scattering in the semi-hard regime  as a superposition of two
different phenomena: color exchanges (which leads to a constant cross
section, that does not grow with $\sqrt{s}$) and the topological
cluster production. The single-pomeron may be caused by the former
component alone, while the double-pomeron may produce a visible cluster
which we will try to associate with the latter.
Let us now compare the dependence of the different kinematical observables
in the UA8 parameterization given above and in our semi-classical formulae.

{\bf The $\phi$-dependence} is explicitly\footnote{Implicitly there
is some dependence due to kinematical limitations: not all values of
the angle between $\vec q_1,\vec q_2$ lead to positive $M_X$.  }
absent in (\ref{eq:pompomdist}).
The same is true in our expressions, if  the total sum of the even and odd
parity combinations is taken.

{\bf The $t_1,t_2$-dependence} factorizes in both cases. Naturally, since
the UA8 expression is written for nucleons while we deal
with partons, their expression has the nucleon form factor and
ours does not. The remaining factor in the UA8 parameterization is

\be
F_{UA8}=&&e^{-q^2\,(b+2\alpha'log(1/\xi))} \nonumber\\
\approx&& e^{-q^2 2.6 GeV^{-2}}
\ee
while we have a square of our basic form-factor $F_{our}=|{\bf J}(q)|^2$.
Although different in shape, they are not so much different in the
range $t=$1-2 GeV$^2$ in which the experiment has been done.
The main scale of the object involved is obviously very close in both cases.
Below we will discuss in more detail the  WA102 data, which have a wider coverage
on the $t$-dependence,

{\bf The $\xi$-dependence.} In the first approximation, putting
$\alpha(0)\approx 1$ into (\ref{eq:pompomdist}), one finds just
$d\xi_1d\xi_2/\xi_1\xi_2 =dy_1dy_2$
where $y_{1,2}$ are Pomeron rapidities, giving a flat rapidity distribution
of a cluster. This contributes ${\rm ln}\,(s)$ to the cross section.
The same dependence is seen in our formulae as well.
In the next approximation one has a correlation between $\xi$ and
the transverse momenta in  (\ref{eq:pompomdist}),  but this was  most probably
never really tested directly in those data, we think.

{\bf The $M_X$-dependence}  is of course the main issue.
The UA8 results are shown in Fig. 2 above. Our qualitative expectations
are a peak at the sphaleron mass, around 3 GeV.
 This is not in contradiction with the data, especially with
the more kinematically constrained ``AND'' set. Unfortunately
the low statistics and   rather crude
resolution of the UA8 experiment in $M_X$ leaves many unanswered questions.
As we have shown in~\cite{jsz02}, at this point we can only calculate the low and
high-$M_X$ parts separately, with the complete treatment in between
still missing. So far we do not have a complete semiclassical prediction for
the exact shape of the $M_X$ dependence, but we are working on it.

{\bf The magnitude} of the double-pomeron cross section
can be estimated qualitatively. One of the reason for that
is that our quark-quark cross section should be extended to
$qg,gg$ collisions and convoluted with appropriate structure
functions. Those should be normalized at the semi-hard scale
corresponding to $-t\sim$ 1-2 GeV$^2$ and the instanton size.
How to do so was discussed in \cite{COS}. In short, each gluon gets an
extra factor of 2 in the cross section relative to a quark (SU(2) color
scaling). The total number of relevant partons was defined there
as the integrated structure functions above $\xi>0.01$, giving about
$N_g\approx4$ gluons and $N_q\approx 4$ quarks
(including the sea and valence quarks), and the elementary
instanton-induced cross section for quark-quark
collision $\sigma_{qq}\sim .017$ mb.
Multiplying all with the extra factor for color singletness 
(double-pomeron) as in (\ref{A19x}), we estimate the total
double-pomeron cross section
for two nucleons to be about  $\sigma_{DD}^{NN}\sim 0.2$ mb,
or about 1 percent of the Pomeron-related part of the cross section.
Parametrically this smallness  comes from the first power of the
instanton diluteness parameter $\kappa=n\rho^4\sim (1/3)^4$ and 
$\alpha_s$ in (\ref{A19x}) for the extra gluon.

The experimental total double-pomeron cross section is
clearly in the same ball-park, although an accurate determination
by extrapolation of the  UA8 data to the whole kinematical region
is too uncertain to quote any specific number.

\section{Parity Even and Odd Clusters}

The  cross section (\ref{A14}) can easily be separated
into the even/odd parity contributions following the natural parity
separation in the form factor (\ref{A20}). The even parity combination is
\be
|{\bf J} (q_{1\perp}) \cdot {\bf J} (q_{2\perp})|^2 =
{\rm cos}^2\phi\,|{\bf J} (q_{1\perp})|^2 \,|{\bf J} (q_{2\perp})|^2
\label{B1}
\ee
while the odd parity combination is
\be
|{\bf J} (q_{1\perp})\times {\bf J} (q_{2\perp}) |^2 =
{\rm sin}^2\phi\,|{\bf J} (q_{1\perp})|^2 \,|{\bf J} (q_{2\perp})|^2
\label{B2}
\ee
where $\phi$ is the azimuthal angle between $(q_{1\perp},q_{2\perp})$.
The double-pomeron cross section for positive and negative parity glue
emission are therefore (per unit rapidity $\eta$)

\be
&&\frac{d\sigma_+}{dQ^2dq_{1\perp}dq_{2\perp}d\eta} = {\rm cos}^2\phi\,
{\bf C}_S\pi\rho^2\,|{\bf J} (q_{1\perp})|^2 \,|{\bf J} (q_{2\perp})|^2
\,\sigma_S(Q)\nonumber\\
&&\frac{d\sigma_-}{dQ^2dq_{1\perp}dq_{2\perp}d\eta} = {\rm sin}^2\phi\,
{\bf C}_S\pi\rho^2\,|{\bf J} (q_{1\perp})|^2 \,|{\bf J} (q_{2\perp})|^2
\,\sigma_S(Q)\,\,.\nonumber\\
\label{B3}
\ee
where ${\bf C}_S$ can also be rewritten as

\be
{\bf C}_S= \frac{3}{N_c^2}\, \alpha_s^2\,{\rm
ln}^2\left(\frac{4\alpha_s}{3\pi}\right)\,
\frac{\Delta (0)}{\kappa}\left(\int dq_\perp{\bf J}^2(q_\perp)\right)^{-2}
\label{B4}
\ee
with $\Delta(0)\approx 0.1$ the `Pomeron' intercept.
The sum of the two, as noted before, is
thus predicted to be independent of $\phi$.

%The present parton-parton cross sections can
%be translated to meson-meson, meson-proton and proton-proton
%diffractive cross sections by multiplying by $4$, $2N_c$
%and $N_c^2$ respectively. Note that at large $\sqrt{s}$ and
%in the CM frame, the  phase space translation is

\be
\frac{d\sigma}{dx\,d\phi\,dt_1\,dt_2} /
\frac{d\sigma}{dQ^2dq_{1\perp}dq_{2\perp}} =\pi\, s \,x
\ee
for symmetric kinematics with

\be
&&q_1=((1-x_1)\sqrt{s}, -q_{1\perp}, (1-x_1)\sqrt{s})\nonumber\\
&&q_2=((1-x_2)\sqrt{s}, -q_{2\perp}, -(1-x_2)\sqrt{s})\,\,,
\ee
$x_1=x_2=1-x$, $t_1=t_2=t$, $|q_{1\perp}|=|q_{2\perp}|=\sqrt{-t}$, and

\be
Q^2=sx^2+2t(1+{\rm cos}\,\phi)\,\,.
\ee

\section{Isosinglet Scalar and Pseudoscalar Production}

The double-pomeron cross section for even/odd parity gluon production
can be readily translated to the isosinglet scalar ($s_0$) and
pseudoscalar ($\eta_0$) $\overline{q}q$ double-pomeron cross sections
through the scale and U(1) anomaly in QCD. In the instanton vacuum
the induced interaction is given by~\cite{kacir}

\be
{\cal S} =&& +
\int\,dz\,\frac 1{2\chi_*}\left(\chi(z) +i\chi_*\sqrt{2N_f}\,\eta_0
(z)\right)^2\nonumber\\
&& +
\int\,dz\,\frac 1{2\sigma^2_*}\left(\sigma(z) +i\sigma^2_*\sqrt{2N_f}\,s_0
(z)\right)^2\,\,,
\label{C1}
\ee
where $\sigma^2_*$ and $\chi_*$ are the compressibility and topological
susceptibility for fermionless QCD.

For small $Q^2$ the sphaleron induced double-pomeron cross section
is mostly mediated by a singular instanton-antiinstanton configuration,
which is about 1 quasiparticle before the unitarity cut. Thus the
mixing (\ref{C1}) amounts to respectively multiplying the gluon
 amplitudes by

\be
&&\sqrt{2N_f}\,\chi_*\,\rho^4\,\nonumber\\
&&\sqrt{2N_f}\,\sigma^2_*\,\rho^4\,\,\,,
\label{C2}
\ee
for the odd/even parities respectively.
So we get from the parity even/odd double-pomeron gluon cross sections
to the parity even/odd double-pomeron isosinglet cross sections by
multiplying the formers with $(\sqrt{2N_f}\,\sigma^2_*\rho^4)^2$ (even) and
$(\sqrt{2N_f}\,\chi_*\rho^4)^2$ (odd). The $\eta_0$ is a mixture of
$\eta',\eta$ with mixing angle $\theta_M\approx 20^0$.

Putting everything together, it follows that the double-pomeron
production of $\eta'$ in $pp$ scattering is given respectively by

\be
\frac 1{N_c^2}\frac{d\sigma_{\eta'}}{dq_{1\perp}dq_{2\perp}d\eta} =
&&\left(\sqrt{2N_f}\,\chi_*\rho^4\,{\rm cos}\,(\theta_M)\right)^2\,{\rm sin}^2\phi\,
\nonumber\\&&\times\,
{\bf C}_S\pi\rho^2\,|{\bf J} (q_{1\perp})|^2 \,|{\bf J} (q_{2\perp})|^2
\,\sigma_S(m_{\eta'}^2)\nonumber\\
%\nonumber\\
%\frac 1{N_c^2}\frac{d\sigma_{\eta}}{dq_{1\perp}dq_{2\perp}d\eta} =
%&&\left(\sqrt{2N_f}\,\chi_*\rho^4\,{\rm sin}\,(\theta_M)\right)^2\,{\rm sin}^2\phi\,
%\nonumber\\&&\times \,
%{\bf C}_S\pi\rho^2\,|{\bf J} (q_{1\perp})|^2 \,|{\bf J} (q_{2\perp})|^2
%\,\sigma_S(m^2_\eta)\,\,,\nonumber\\
\label{C3}
\ee
while the double-pomeron production of heavy isosinglet scalars reads

\be
\frac 1{N_c^2}\frac{d\sigma_{s}}{dq_{1\perp}dq_{2\perp}d\eta} =&&
\left(\sqrt{2N_f}\,\sigma^2_*\rho^4\right)^2\,{\rm cos}^2\phi\,
\nonumber\\&&\times\,
{\bf C}_S\pi\rho^2\,|{\bf J} (q_{1\perp})|^2 \,|{\bf J} (q_{2\perp})|^2
\,\sigma_S(m_s^2)\,\,.\nonumber\\
\label{C4}
\ee

\section{Exclusive Double-Pomeron: WA102 }

The most natural general prediction which follows from our approach is
that since the inclusive clusters are predicted to have a mass
of about 3 GeV, the exclusive glueball states -- such as the 
scalar at 1.7 GeV and pseudoscalar above 2 GeV --
can be significant. As the latter has not yet been identified experimentally,
we focused instead on  the $\eta'$, known to interact strongly with the
glue.

{\bf The $\phi$ dependence} is the
most dazzling empirical feature of WA102.
The data exhibit different azimuthal dependence
of the partial cross section for various $J^{PC}$ quantum numbers of the
final states. In Fig.~\ref{fig:phi} we show the dependence on $\phi$ for
two $0^{-+}$ channels, $\eta,\eta'$, which is clearly
${\rm sin}^2(\phi)$. This is in complete agreement with the dependence
expected from our calculations. On the other hand for the scalar
glueball the WA102 finds a distribution with a maximum at $\phi=0$ (not
shown here). In this case, we predict ${\rm cos}^2(\phi)$.

\begin{figure}
\begin{center}
\mbox{\epsfig{file=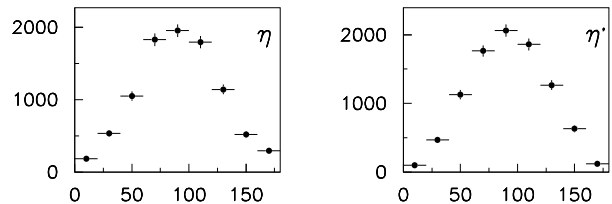,width=8cm}}
\end{center}
\caption[]{
Dependence of the double-pomeron cross section on the
azimuthal angle $\phi$ for $0^{-+}$
production in the final states from Ref.~\cite{kirk}}
\label{fig:phi}
\end{figure}

The {\bf $t$-dependence} is also very important to compare.
As shown in our previous paper \cite{sz01}, the form-factor 
basically represents the size and the shape of the instantons,
and although the latter is established rather well with $\rho\approx 0.3$ fm
\cite{SS_98}, the shape is not. So identifying  the observed $t$ dependence
with our formfactor one can learn about the actual shape of the instantons in the QCD vacuum.

The measured dependence on the transverse momentum $|t|$ is shown in
Fig.~\ref{fig:dft}. It is best parametrized as~\cite{kirk}
$\beta\,|t|^n\,e^{-b\,|t|}$ with $n\approx 1-2$,
$10^3\beta\approx 110$ and $b\approx 11\,$GeV$^{-2}$.
This dependence in our case is carried by the ${\bf
F}^4$ factor in the differential cross section. The dependence of ${\bf
F}$ on $\sqrt{-t}$ in units of the instanton size is shown in
Fig.~\ref{fig:ff} for ${\bf a}=0,1/4,1/2,3/4$. Clearly, the empirical
results are in agreement with a non-vanishing ${\bf a}$ as originally
suggested in~\cite{sz01}, and in disagreement with the Pomeron
expectation. For large $|t|$ the sphaleron induced form factor falls
as $e^{-6\,\rho\sqrt{-t}}$ with $6\,\rho\approx 7$ GeV$^{-1}$.

\begin{figure}
\begin{center}
\mbox{\epsfig{file=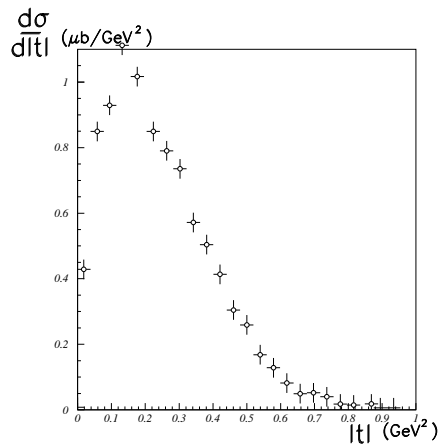,width=6cm}}
\end{center}
\caption[]{
The WA102 measured dependence of the double-pomeron cross section for
$\eta'$ production on the momentum transfer $|t|$~\cite{koc}.}
\label{fig:dft}
\end{figure}

The next Fig.\ref{fig:ff} compares this experimental $t$-dependence
from $(d\sigma/dt)^{1/4}$
to the instanton formfactor. The 4 theoretical curves are from our
previous paper  \cite{sz01}. They show the predicted shape for
4 values of the instanton tail modification parameter
$a$. The experimental points show the same shape as predicted.
Furthermore, they are right in the middle of the $a$ interval assumed, 
so that $a\approx 1/3$ would fit very well.
Note that the relative errors (which are not shown)
are large for points on the right side of the experimental quotes.

\begin{figure}
\begin{center}
\mbox{\epsfig{file= 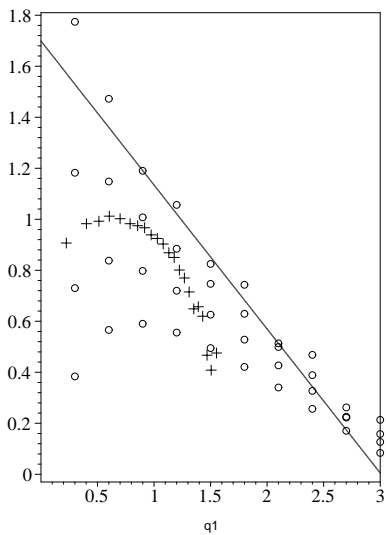,width=5cm}}
\end{center}
\caption[]{
The log of the form factor $|{\bf J}(q_\perp)|$ versus $q1=\rho\sqrt{-t}$
where $\rho$ is the instanton size. The 4 
theoretical curves are for the instanton shape parameter
${\bf a}=0,1/4,1/2,3/4$ (top to
bottom)~\cite{sz01}. The crosses show the data points (without error bars)
of the data shown in the previous figure, as $(d\sigma/dt)^{1/4}$
with the same units of $q1$ (arbitrary normalization).}
\label{fig:ff}
\end{figure}

An estimate of the $\eta'$ cross section  in $pp$ double-pomeron
follows from our projected result (\ref{C3}), i.e.

\be
\frac{\sigma (\eta')}{\sigma_{IN}} \approx (0.125)\times (6\,10^{-4})\times (10^{-1})\,\,.
\label{C5}
\ee
where the first factor is due to the singlet projection, the second factor
to the $\eta'$ projection and the third factor to the fact that the
$\eta'$ mass is lower than the sphaleron mass which is an extra penalty
in the partial cross section. For $\sigma_{IN}\approx 30$ mb at
$\sqrt{s}\approx 30$ GeV, we predict $\sigma (\eta')\approx 225$ nb in
comparison to the $588$ nb observed empirically~\cite{kirk}.

\section{Conclusions}

We have addressed the issue of inclusive and exclusive double-pomeron
 scattering in the semi-hard regime, using the semiclassical
theory of high energy collisions recently developed.
We have found that it works remarkably well, explaining even such
details as correlation between the azimuthal distributions and
quantum numbers of the cluster.

We have shown that the semiclassical double-pomeron cross section relates
simply to the semiclassical inelastic cross section: overall it is about $10\%$
of the latter due to an extra gluon and extra color restrictions.
This corresponds to a large cross section in absolute magnitude, well
above the Pomeron factorization, as the UA8 experiment indeed
observed. We also have shown that all the distributions
of the UA8 inclusive data over 6-d kinematical space is quite
compatible with our predictions.

The exclusive results from WA102 experiment seem to confirm this
theory even more. We have a very good agreement for scalar glueball
and $\eta'$ productions, which show very different azimuthal dependence.

As example of a disagreement let us mention that in our theory
central double-pomeron semiclassical production of $\eta$
is found to be suppressed by almost 2 orders of magnitude in
comparison to $\eta'$: a suppression of $10^{-1}$ is due to the mixing
angle, and an extra suppression of $10^{-1}$ is due to its lower mass
in the partial cross section. This strong suppression is not observed.
We think that all light hadrons such as  $\eta$ and
$\pi^0$ and possibly $f_0(600)$ are too far from the sphaleron mass
scale of about 3 GeV to be reliably calculated in the same manner.
We currently suspect their production to rely on a different mechanism, 
and will report on it elsewhere.

Since some sense has been made of the single- and double-pomeron
physics in the context of semiclassical 
dynamics of the gauge fields, related with tunneling and topoly,
 new rounds of experiments 
seem to be justified in this context more than ever. We believe that the
RHIC detectors and especially STAR can do a lot in the $pp$ mode, 
with and without diffraction. Clearly they have a potential
to clarify further the nature and characteristics of
the central production in the semi-hard regime.

\vskip 1.25cm
{\bf Acknowledgments}
\\\\
This work was supported in parts by the US-DOE grant
DE-FG-88ER40388. One of us I.Z. would like to thank V. Vento for
discussions and La Comisi\'on de Intercambio Cultural, Educativo y
Cient\'{\i}fico entre Espa\~na y Estados Unidos de Am\'erica for
supporting his visit to the University of Valencia where this work
was initiated. We both thank P.Schlein for his comments on the manuscript
and G. Papp for help with some of the figures.

%\appendix

\end{narrowtext}
\end{document}